\documentclass[12pt]{iopart}   
\usepackage{graphicx}
\usepackage{epsfig}
\begin{document}

\title[Dynamics at barriers in bidirectional two-lane exclusion processes]{Dynamics at barriers in bidirectional two-lane exclusion processes}

\author{R\'obert Juh\'asz} 
\address{Research Institute for Solid
State Physics and Optics, H-1525 Budapest, P.O.Box 49, Hungary}
\ead{juhasz@szfki.hu} 

\begin{abstract}
A two-lane exclusion process is studied where particles move in the
two lanes in opposite directions and are able to change lanes. 
The focus is on the steady state behavior in situations where a positive 
current is
constrained to an extended subsystem 
(either by appropriate boundary conditions 
or by the embedding environment) where, in the absence of the constraint, 
the current would be negative.  
We have found two qualitatively different types of steady states and 
formulated the conditions of them in terms of the transition rates.   
In the first type of steady state, a localized cluster of particles forms with
an anti-shock located in the subsystem and the current vanishes exponentially
with the extension of the subsystem. This behavior is analogous to that of the
one-lane partially asymmetric simple exclusion process, and can be realized 
e.g. when the local drive is induced by making the jump rates in two lanes
unequal.    
In the second type of steady state, which is realized e.g. if the local 
drive is induced purely by the bias in the lane change rates, and which has
thus no counterpart in the one-lane model, 
a delocalized cluster of particles forms 
which performs a diffusive motion as a whole and, as a consequence, the 
current vanishes inversely proportionally to the extension of the
subsystem.
The model is also studied in the presence of quenched disordered, 
where, in case of delocalization, 
phenomenological considerations predict anomalously
slow, logarithmic decay of the current with the system size in contrast with the usual power-law.  
\end{abstract}

\pacs{05.70.Ln, 87.16.aj, 87.16.Uv}
\maketitle

\newcommand{\bc}{\begin{center}}
\newcommand{\ec}{\end{center}}
\newcommand{\be}{\begin{equation}}
\newcommand{\ee}{\end{equation}}
\newcommand{\beqn}{\begin{eqnarray}}
\newcommand{\eeqn}{\end{eqnarray}}

\section{Introduction}

Transport in cells is an intensively studied field of cell biology as it is 
important for cellular organization and function, moreover, malfunction of
transport processes may lead to severe diseases \cite{howard,sw,gross}. 
Intracellular transport is realized by a class of proteins called molecular
motors which are able to convert chemical energy released form hydrolysis of
ATP into mechanical work and thereby perform directed movement on the filament
network and carry different cargoes from one part of the cell to the other \cite{sw}.  
Although basic features of this system are well known there are still 
important details which have not cleared up yet and, for instance, 
concurring models of bidirectional transport exist currently \cite{gross}.

This important issue has motivated much theoretical work on special 
driven lattice gas models which have been used as minimal models of
cooperative behavior of molecular motors \cite{schad}. 
The cornerstone of these investigations is the asymmetric simple
exclusion process (ASEP) \cite{mcdonald,spitzer} in which   
particles sitting on a one-dimensional lattice jump stochastically to
a neighboring site provided the latter is empty. 
Due to its simple formulation, this process has become a 
paradigmatic model of interacting many-particle systems 
which have nonequilibrium steady-states \cite{liggett,zia,schutzreview}.
The system, when coupled to particle reservoirs at the ends, shows boundary 
induced phase transitions \cite{krug} and the phase diagram
is exactly known \cite{derrida,schutzdomany,blythe}.  

The ASEP has been generalized in many different directions 
in order to take into account more details of real transport systems. 
Such generalized models are the two or multilane models which consist of two
or more ASEPs coupled in parallel, where particles are 
able to jump to other lanes.   
Formally, on the grounds of the directionality of the motion in the lanes, these models can be divided in three classes. 
First, particles move totally asymmetrically (unidirectionally) 
within lanes and the direction of motion 
is the same in all lanes \cite{pk,mitsudo,harris,reichenbach,jiang,cai}.
This generalization is inspired by the circumstance that the microtubule
filament that serves as a track for certain molecular motors is not strictly
one-dimensional but consists of many proto-filaments.  
Such models have also been investigated in the context of vehicular
traffic on multi-lane roads; for a review, see e.g. Ref. \cite{santen}.   
Second, two-lane models with unidirectional motion in one lane and symmetric
(diffusive) motion in the other one have also been 
introduced \cite{tsekouras,es}.  
These models have been intended to take into consideration that real molecular
motors can unbind from the filament, diffuse in the surrounding fluid and
rebind again to the filament. The unbinding and rebinding has also been taken
into account by simpler models where the ASEP is extended by creation and
annihilation of particles \cite{frey}, as well as by more realistic models
where the ASEP is coupled to a finite compartment in which the motion is
diffusive \cite{nieuwenhuizen,klumpp,muller,tailleur}.   
Third, a two-lane model with unidirectional motion in the lanes but with
opposite orientation of the lanes has also been studied \cite{j1,kl}. 
This arrangement can be thought of as a simplified model of motors moving 
along two oppositely oriented filament tracks which are located in a tube-like 
narrow compartment.   
The motion of a single particle in this environment has been found to have 
an enhanced diffusion coefficient compared to that of the symmetric random
walk \cite{kl}. This type of active diffusion can be realized experimentally
and it is also present {\it in vivo} \cite{gross}. 
An alternative interpretation of this model is that a single filament is placed
in a narrow compartment where there is a steady convection of the medium 
in the opposite direction compared to the polarity of the filament.

The theoretical studies have concluded that, in all three cases, the
steady-state behavior is richer than that of the one-lane ASEP,
furthermore, interesting phenomena take place in such systems, 
for instance, synchronization or localization of density shocks. 
The latter can be observed experimentally in the 
traffic of kinesin motors in accordance with theoretical 
predictions \cite{konzack,nishinari,greulich}.    
    
This work wishes to contribute to the studies on the less 
examined bidirectional models of the third class.
Earlier studies have concentrated either on the one particle dynamics
\cite{kl} or on the case of weak coupling between the channels, which can be
treated by means of a hydrodynamic description \cite{j1}. 
In the present work, the case will be considered 
where the inter-lane jump rates are comparable with the intra-lane jump rates. 
A special property of this model is that, under certain conditions, 
the stationary current can be reversed when the global density of
particles is varied. In other words, if particles are present in the system 
with a sufficiently high density they are able to flow in the 
opposite direction as a single particle.  
We shall give the condition of this phenomenon in terms of the transition
rates of the process. 
Subsequently, we shall analyze the quantitative consequences of this property 
in arrangements where the system contains a finite region where the local bias is 
opposite to the global one. 
It will turn out that the steady state at such an inclusion with reverse
bias, or ``barrier'' is qualitatively different from that of the analogous 
one-lane partially asymmetric simple exclusion process in the 
reverse bias regime. 
In the latter model, a shock forms in the middle of the barrier 
and the stationary current tends to zero exponentially with the size of the
barrier \cite{sandow,ramaswamy,blythe}.  
As opposed to this, in the two-lane model to be studied, a shock still appears
but, under certain conditions, it may be delocalized, 
which leads to that the current vanishes much slower:
it is inversely proportional to the size of the barrier. 
A direct consequence of this slow mode is that, in case of quenched disordered
lane-change rates, the current $J$ vanishes with the size of the system $L$ 
as $J(L)\sim (\ln L)^{-1}$ in the presence of a global bias, 
which is slower than the usual 
power-law decay characteristic of disordered partially asymmetric one-lane 
models \cite{tripathy,krug2,jsi,barma,j2}.  

The paper is organized as follows. 
The model will be defined in Sec. 2. The process in the presence of a single
particle is studied in
Sec. 3. The periodic system and the phases determined by the bulk rates are
discussed in Sec. 4. In Sec. 5, the phase diagram of the open system is
determined. Sec. 6 is devoted to heterogeneous models including one-barrier
systems and random disorder. Some calculations are presented in the Appendix
and the results are summarized in Sec. 7. 

\section{The model} 

\begin{figure}[h]
\includegraphics[width=0.6\linewidth]{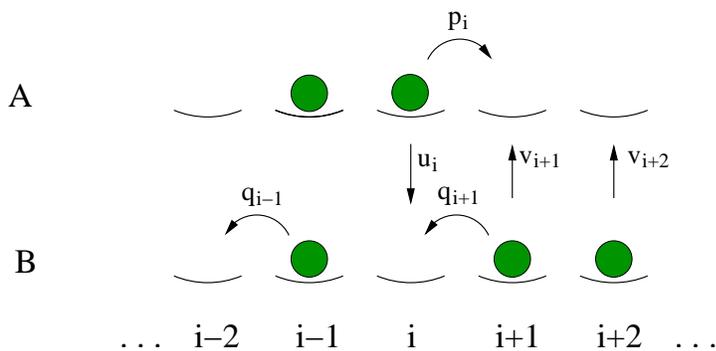}
\caption{\label{fig1} Schematic view of the model in the bulk. Arrows indicate
  the allowed transitions.}
\end{figure}

The mode is defined on two parallel one dimensional
lattices with $L$ sites, denoted by $A$ and $B$, the sites of which 
are either empty or occupied by a particle. 
We shall study periodic chains (or rings), where site $1$ and $L$ are 
neighbors, as well as open chains.  
The configuration of the system is specified by the set of
occupation numbers $n^{A,B}_l$, $l=1,2,\dots,L$, which are zero (one) for empty
(occupied) sites. 
On the space of configurations, a continuous-time Markov process is defined 
by the rates of allowed transitions, are the followings    
for periodic chains. 
On site $l$ of lane $A$($B$), a particle attempts to jump to the adjacent
site on its right(left) hand side with rate $p_l$($q_l$) and the attempt is
successful if the target site is empty.   
Furthermore, a particle residing at site $l$ of chain $A$($B$) jumps 
to site $l$ of chain $B$($A$) with rate $u_l$ ($v_l$), 
provided the target site is empty.  
The model is illustrated in Fig. \ref{fig1}. 

\begin{figure}[h]
\includegraphics[width=0.6\linewidth]{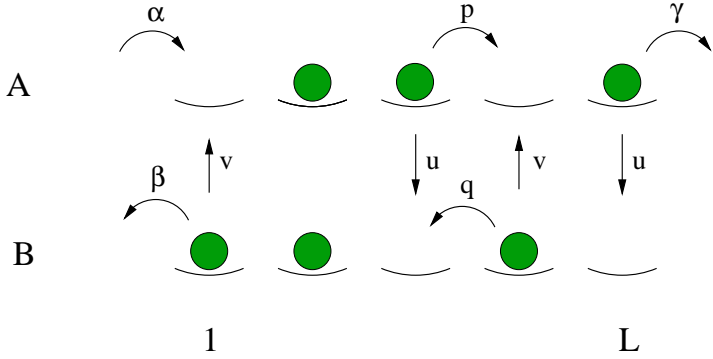}
\caption{\label{fig2} Schematic view of the open system. Arrows indicate the
  allowed transitions.}
\end{figure}

For open systems, the above process is modified at surface sites 
$1$ and $L$ as follows. 
On the first site of lane $A$ particles are injected with rate $\alpha$, 
provided this site is empty. 
On site $L$ of lane $A$ particles are deleted with rate $\gamma$, whereas 
on the first site of lane $B$ they are deleted with rate $\beta$. 
In case of open systems, we restrict ourselves to a homogeneous bulk, 
i.e. $p_l=p$, $q_l=q$, $u_l=u$, $v_l=v$, see Fig. \ref{fig2}.
Moreover, we shall focus on the reverse bias regime, 
where a single particle is driven to the left in the bulk 
and are mainly interested in under which conditions a positive current
can be forced to the system by injecting particles at site $1$.
For sake of simplicity, particles are not injected at site $L$ and the exit
rate $\gamma$ at that site is set to $1$.  


\section{One-particle dynamics}

\begin{figure}[h]
\includegraphics[width=0.4\linewidth]{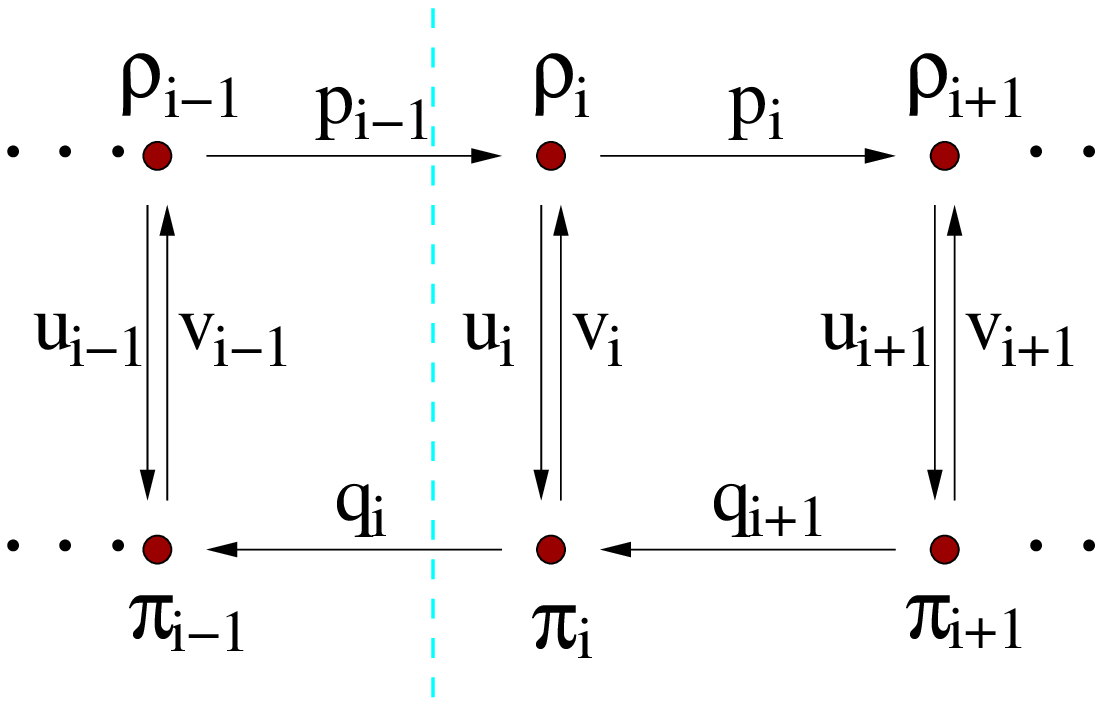}
\includegraphics[width=0.4\linewidth]{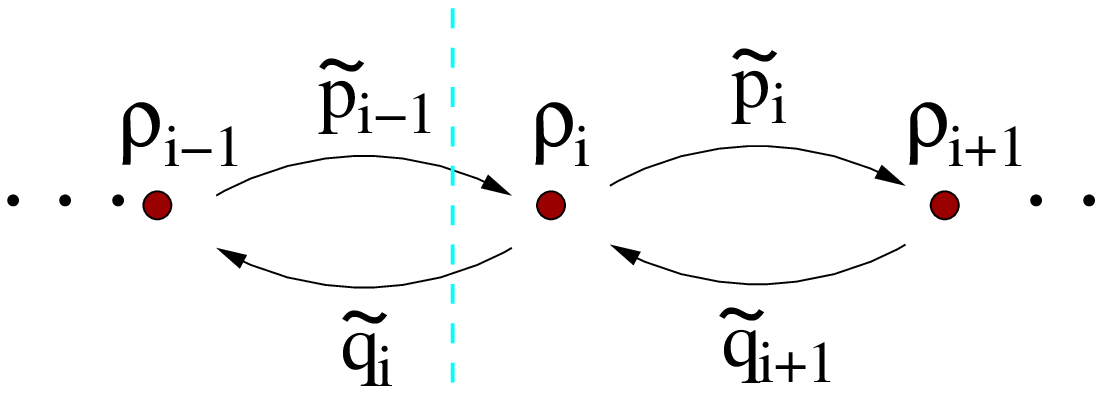}
\caption{\label{fig3} Left: stationary probabilities on a finite ring with a 
single particle. Right: partially asymmetric one-lane model obtained by the
  elimination of lane $B$. The stationary current across the dashed line is denoted by $J$.}
\end{figure}

Let us examine the motion of a single particle in the system, a situation
which will be relevant for the analysis of the stability of shocks. 
Although, detailed-balance does not hold in this system as the motion of 
the particle is unidirectional within the lanes, we shall show that 
an effective potential can still be defined. 
Let us consider the steady state of a particle on a finite ring. 
The probability that the particle is found at site $i$ of lane $A$($B$)
is denoted by $\rho_i$($\pi_i$). The probability current $J$ 
flowing parallel to the lanes (see Fig. \ref{fig3}) 
can be written as 
\be
J=p_{i-1}\rho_{i-1}-q_{i}\pi_{i}. 
\label{J2}
\ee 
Conservation of probability at site $i$ of lane $A$ and $B$ imply
\beqn 
p_{i-1}\rho_{i-1} + v_i\pi_i -(p_i+u_i)\rho_i = 0 
\label{con1}
\\
q_{i+1}\pi_{i+1} + u_i\rho_i -(q_i+v_i)\pi_i  = 0.
\label{con2}
\eeqn
Let us assume that $v_i>0$ for $i=1,2,\dots,L$. 
Expressing $\pi_i$ from Eq. (\ref{con1}) and substituting into Eq. (\ref{J2})
we obtain
\be 
J=\tilde p_{i-1}\rho_{i-1}-\tilde q_{i}\rho_{i},
\ee
where 
$\tilde p_{i-1}=p_{i-1}(q_i+v_i)/v_i$ and $\tilde q_{i}=q_{i}(p_i+u_i)/v_i$.
Formally, this equation is identical to that describing the stationary current
in a one-lane partially asymmetric model with forward and backward rates 
$\tilde p_{i}$ and  $\tilde q_{i}$, see Fig. \ref{fig3}. 
This correspondence with a one-lane model is not perfect as the
occupations $\rho_i$ alone are not normalized to one, i.e. $\sum_{i=1}^L\rho_i<\sum_{i=1}^L(\rho_i+\pi_i)=1$.   
Nevertheless, the current differs from that of the equivalent one-lane
model only by a positive $O(1)$ factor. 
This loose formulation of the correspondence will
be sufficient for latter reasoning as the only information we need is the sign
of the current. 
For the equivalent one-lane model a potential $U^A_i$ at site
$i$ can be defined by $\Delta U_i\equiv U^A_i-U^A_{i-1}=\ln (\tilde
q_i/\tilde p_{i-1})$. 
This reads in terms of the rates of the two-lane model: 
\be 
\Delta U_i\equiv U^A_i-U^A_{i-1}=\ln \frac{q_i(p_i+u_i)}{p_{i-1}(q_i+v_i)}.
\label{U}
\ee 
In fact, the potential itself exists even in the case when not all $v_i$ are
positive. 
This effective potential can be obtained in a more direct way by 
considering the
steady state of a single particle in a closed system, i.e. open chain with 
$\alpha=\beta=\gamma=0$. In this case $J=0$ and  Eqs. (\ref{J2}-\ref{con2})
can be equivalently written as:
\beqn 
\rho_{i-1}/\rho_{i}=\frac{q_i(p_i+u_i)}{p_{i-1}(q_i+v_i)} \\
\pi_{i}/\pi_{i+1}=\frac{q_{i+1}(p_i+u_i)}{p_{i}(q_i+v_i)} \\
\pi_i/\rho_i=\frac{p_i+u_i}{q_i+v_i}.
\eeqn
From these relations, one can see that the steady state is identical to that 
of an equilibrium system, 
where detailed balance is satisfied and the potential 
is given by Eq. (\ref{U}) in lane $A$ and by 
$U^B_i-U^A_i=\ln [(p_i+u_i)/(q_i+v_i)]$ in lane $B$.
We thus conclude that the motion of a single particle can be effectively 
described by a one-dimensional motion in the potential $U^A_i$.   

Next, we examine the dynamics of a single vacancy in the system when all 
but one of the sites are occupied.    
It is easy to see that the dynamics of a lonely vacancy differs from that of a
lonely particle in that the direction of transitions (i.e. the arrows in
Fig. \ref{fig3}) are reversed. This is equivalent with a single particle
dynamics analyzed above, however, in a modified environment where the rates 
$p_{i-1}$ and $q_i$ of the original environment are interchanged for all $i$. 
The effective potential $\overline{U}^B_i$ associated with a single vacancy is thus given by  
\be 
\Delta \overline{U}_i\equiv \overline{U}^B_i-\overline{U}^B_{i-1}=\ln \frac{p_{i-1}(q_{i+1}+u_i)}{q_{i}(p_{i-1}+v_i)}.
\label{Ubar}
\ee 
In a system which is homogeneous in the bulk a single particle is
driven from left to right if $\Delta U<0$, whereas it is driven from
right to left if $\Delta U>0$. In the latter case we speak of 
{\it reverse bias} and in the followings, we shall focus on this regime. 
The condition $\Delta U>0$ of reverse bias in the bulk reads in 
terms of jump rates as : 
\be
q/p>v/u.
\label{rev}
\ee
Note that the potential difference experienced by a vacancy can
have any sign in the reverse bias regime. Namely, it is driven to the
right if $\Delta\overline{U}<0$, or equivalently if $q/p>u/v$
while it is driven to the left if $q/p<u/v$.  
This contrasts with the one-lane partially asymmetric exclusion process, where 
$\Delta U_i=-\Delta \overline{U}_i$ always holds.

\section{Periodic system}

\begin{figure}[h]
\includegraphics[width=0.5\linewidth]{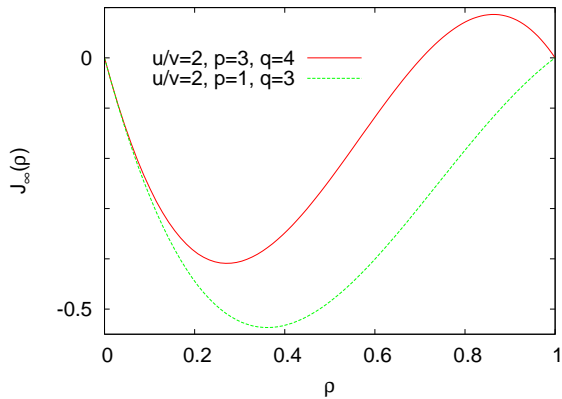}
\caption{\label{fig4} The current as a function of the particle density
 $\varrho$ for two different set of bulk rates.}
\end{figure}

Let us consider a homogeneous, periodic system with $2L$ sites and 
$N$ particles. 
It is straightforward to check that the probability of configurations 
$\{n_i^A,n_i^B\}$ is of factorized form in the steady state: 
\be 
P(\{n_i^A,n_i^B\})=
Z^{-1}\left[\prod_{i=1}^L\rho^{n_i^A}(1-\rho)^{1-n_i^A}\pi^{n_i^B}(1-\pi)^{1-n_i^B}\right]
\delta\left(\sum_{i=1}^{L}(n_i^A+n_i^A),N\right),
\label{fact}
\ee
where $\delta(i,j)$ is the Kronecker symbol and
the factor $Z\equiv\sum_{\{n_i^A,n_i^B\}}P(\{n_i^A,n_i^B\})\delta(\sum_{i=1}^{L}(n_i^A+n_i^A),N)$
ensures normalization.   
The parameters $\rho$ and $\pi$ in Eq. (\ref{fact}) lie in the 
interval $(0,1)$ and satisfy the relation 
\be
u\rho(1-\pi)=v\pi(1-\rho).
\label{rhopi}
\ee
Keeping the density of particles $\varrho\equiv N/(2L)$ constant 
and performing the thermodynamic limit $L\to\infty$,  
the factorized form in Eq. (\ref{fact}) results in a simple form of the current:
$J_{\infty}=p\rho(1-\rho)-q\pi(1-\pi)$. Here, $\rho$ and $\pi$ are fixed by 
the density through $\varrho=(\rho+\pi)/2$ and are interpreted 
as the particle densities in lane $A$ and $B$, respectively. 
The current, as a function of $\rho$, reads as  
\be
J_{\infty}(\rho)=\rho(1-\rho)\left[p-\frac{qv/u}{\rho+(v/u)(1-\rho)^2}\right].
\label{fund}
\ee
The fundamental diagram of the model, i.e. the current plotted against the
density $\varrho$, can be seen in Fig. \ref{fig4}. 
Depending on the bulk rates, one can distinguish between two domains of the
parameter space in the reverse bias regime, which are characterized by
qualitatively different shape of the curve $J(\varrho)$. 
It is straightforward to show that if 
\be
q/p>u/v,
\label{SB}
\ee
the fundamental diagram has a single extremum (minimum) and the 
current is negative for any density $0<\varrho<1$. 
If condition (\ref{SB}) holds, we shall speak of a {\it strong bias} (SB). 
Note that inequalities (\ref{rev}) and (\ref{SB}) imply that $q>p$
is a necessary condition of strong bias. 
If, however, 
\be
q/p<u/v
\label{WB}
\ee 
holds, there exists a non-trivial density $\varrho_0$ in the interval 
$(0,1)$, at which the current is zero, i.e. $J_{\infty}(\varrho_0)=0$. 
The corresponding densities in lane $A$ and $B$ can be easily calculated: 
\be
\rho_0=
\left(\sqrt{\frac{qv}{pu}}-\frac{v}{u}\right)\left(1-\frac{v}{u}\right)^{-1},
\qquad 
\pi_0=\sqrt{\frac{pu}{qv}}\rho_0.
\label{rhostar}
\ee
In this case, the 
fundamental diagram has two extrema (a maximum and a minimum) and, 
for high enough densities ($\varrho_0<\varrho<1$), the current is positive.   
If inequality (\ref{WB}) is satisfied, we shall speak of a 
{\it weak bias} (WB). Inequalities (\ref{rev}) and (\ref{WB}) imply that the
necessary condition of weak bias is $v<u$. 
Recalling the dynamics of a single vacancy discussed in the previous section,
we can see that, in the strong bias regime, the vacancy is driven from left
to right ($\Delta \overline{U}<0$), whereas, in the weak bias regime,
it is driven from right to left ($\Delta \overline{U}>0$).   
The connection between the direction of motion of a single vacancy and
the shape of the fundamental diagram can be immediately seen. 
If the vacancy moves from left to right ($\Delta \overline{U}<0$), 
the current is negative for finite $L$ and $N=2L-1$. 
Therefore the slope of the curve $J_{\infty}(\varrho)$ must be negative at
$\varrho=1$, i.e. $\lim_{\varrho\to 1}\frac{d J_{\infty}(\varrho)}{d\varrho}<0$. 
If the vacancy moves from right to left ($\Delta \overline{U}>0$), 
the current is positive and      
 $\lim_{\varrho\to 1}\frac{d J_{\infty}(\varrho)}{d\varrho}>0$.

In the borderline case $q/p=u/v$, the motion of a vacancy is diffusive
($\Delta \overline{U}=0$) and
$\lim_{\varrho\to 1}\frac{d J_{\infty}(\varrho)}{d\varrho}=0$. Apart from this anomaly
at $\varrho=1$, the shape of the fundamental diagram is similar to that in the
strong bias regime.

\section{Open system}
\label{os}

In this section, we shall investigate the steady state of an open system in
the reverse bias regime, i.e. when the bulk rates satisfy 
inequality (\ref{rev}). 

\begin{figure}[h]
\includegraphics[width=0.4\linewidth]{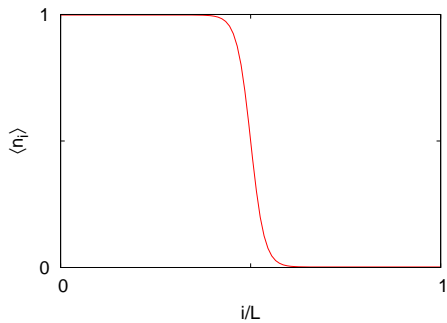}
\caption{\label{fig5} Schematic density profile of the one-lane partially
  asymmetric simple exclusion process in the reverse bias regime $q>p$.}
\end{figure}

First, we recall the steady state of the one-lane 
partially asymmetric simple exclusion process 
in the reverse bias regime ($q>p$) \cite{sandow,ramaswamy,blythe}.
For finite injection rate and zero exit rate at site $1$, 
an anti-shock (or domain wall) develops in 
the density profile, where the density changes rapidly with
the position (see Fig. \ref{fig5}). 
As a consequence of particle-hole symmetry of this model, the anti-shock is
located in the middle of the system. With the distance measured from the
middle of the system, the density tends exponentially to one(zero) 
on the left(right) hand
side of the anti-shock, and the shock region, where the density differs
significantly from the bulk value has a characteristic width 
$\xi\sim 1/\Delta U= 1/\ln(q/p)$. 
The stationary current decreases with the system size as 
$J(L)\sim e^{-L\Delta U/2}$ and tends to zero 
in the limit $L\to\infty$ \cite{blythe}. 

Returning to the two lane model, let us investigate the possibility of an 
anti-shock type steady state.
In the low density domain (i.e. on the right hand side of the domain wall) and
far away from the shock region, particles meet each other very rarely.
Therefore, they can be regarded as freely moving particles. 
The same is true for vacancies in the high density
domain (i.e. on the left hand side of the anti-shock). 
Within the frame of this phenomenological description, 
we can formulate necessary conditions of the stability of the domain wall. 
First, particles which have come off from the high density domain and
penetrated into the low density domain must be driven back to the high density
domain. This condition is equivalent to 
$\Delta U>0$ or, in terms of jump rates, to inequality (\ref{rev}), so
it is always satisfied in the reverse bias regime. 
Second, vacancies which penetrate into the high density domain must be driven
back to the low density domain. This condition is equivalent to 
$\Delta \overline{U}<0$ or, in terms of jump rates, to inequality (\ref{SB}). 
We thus conclude that an anti-shock type steady state in the
reverse bias regime is possible only in the case of a strong bias.  
This analysis shows that the steady state of the model for strong bias and
for weak bias are much different; therefore the two cases will be
discussed separately. 

\subsection{Strong bias}

First, we assume that the bias is strong, i.e. the bulk
rates satisfy inequality (\ref{SB}). 
The exit rate $\gamma$ is set to $1$ and we are interested in the
steady state for given boundary rates $\alpha>0$ and $\beta$. 
Although, we could not find the exact form of the steady state, the
shape of the density profile on a macroscopic scale and the leading
size dependence of the current can be obtained by means of phenomenological
arguments. 
As the localized domain wall is stable for strong bias 
we expect the density profile to be qualitatively similar to that 
of the one-lane partially asymmetric simple exclusion process. 
For $\beta=0$ and for any $\alpha>0$, an anti-shock develops in the interior
of the system, which separates a high density domain from a low
density domain. The center of the shock region is located at some  
$0<l^*<L$ and the density tends exponentially to zero(one) in the
low(high) density domain on a characteristic scale 
$\xi\sim 1/\Delta U$ ($\overline{\xi}\sim 1/|\Delta\overline{U}|$).
The position of the domain wall $l^*$ is determined by the condition
that, at stationarity, the traveling time of particles from the shock region
to site $L$ must equal to the traveling time of vacancies through the
high density domain to site $1$, otherwise the domain wall would move
with a finite velocity. 
Thus, for $\xi,\overline{\xi}\ll L$, we may write 
\be
e^{l^*|\Delta\overline{U}|}\sim e^{(L-l^*)\Delta U}.
\label{dwstat}
\ee
Introducing the fraction $r$ of the system which is 
occupied by the high density domain 
in the limit $L\to\infty$, i.e. $r\equiv\lim_{L\to\infty}l^*(L)/L$, 
Eq. (\ref{dwstat}) yields: 
\be 
r=\left(1+|\Delta\overline{U}|/\Delta U\right)^{-1}=
\left(1+\ln\frac{q(p+v)}{p(q+u)}/\ln\frac{q(p+u)}{p(q+v)}\right)^{-1}.
\label{r}
\ee
As can be seen,
unlike for the one-lane partially asymmetric simple exclusion process,
$r$ differs from $1/2$ if $u\neq v$, 
which is the consequence of the breaking of the particle-hole symmetry.
For large $L$, the stationary current, which is proportional to the inverse
of traveling time, tends exponentially to zero with increasing system size:
\be 
J(L)\sim e^{-\Delta U(1-r)L}=
\left(\frac{q(p+v)}{p(q+u)}\right)^{(1-r)L}.
\label{JSB}
\ee

If $\beta>0$ (and $\alpha>0$), the high density domain shrinks to the left
boundary of the system; in other words, the low density domain extends over the
entire system and a boundary layer forms at the left
boundary. The width of the boundary layer increases with decreasing $\beta$ 
but remains finite for finite $\beta$, which means that $r=0$.  
The expression of the current in Eq. (\ref{JSB}) is valid with $r=0$. 

\subsection{Marginal bias}

Next, the marginal case is considered, when $q/p=u/v$ or, equivalently, 
$\Delta\overline{U}=0$. 
From Eq. (\ref{r}) we obtain $r=1$. Therefore, if $\beta=0$ and $\alpha>0$ the
high density domain extends over the whole system and a boundary layer forms
at the right boundary. 
The characteristic scale $\overline{\xi}=1/|\Delta\overline{U}|$ diverges, 
therefore the density must
tend slower than exponentially to the bulk value. 
The motion of a single vacancy is diffusive, so, far from the right boundary,
where the density of vacancies is close to zero, 
they can be approximatively regarded as independent, diffusing particles. 
When a system of independent, diffusing particles 
is coupled to boundary reservoirs with
different densities, the steady state density profile interpolates linearly
between the densities of reservoirs. 
Therefore we conclude that, for sites far away from the right boundary, the
profile is linear: 
$1-\langle n^A_i\rangle\sim i/L$, $1-\langle n^B_i\rangle\sim i/L$, $i\ll L$.    
It follows then that the current is inversely  proportional to the system size:
\be
J(L)=\alpha(1-\langle n^A_1\rangle)\sim L^{-1}.
\ee

For $\beta>0$ (and $\alpha>0$), the high density domain shrinks to the left
boundary ($r=0$). For a given $\beta$, the width of the boundary layer is 
significantly larger than in the case of strong bias since the profile 
in the high density domain approaches the bulk value much slower.    
The current decreases exponentially as given in Eq. (\ref{JSB}) with $r=0$.

Numerical simulations are in agreement with the above results
obtained for strong and marginal bias. 

\subsection{Weak bias}
\label{oswb}

\begin{figure}[h]
\includegraphics[width=0.5\linewidth]{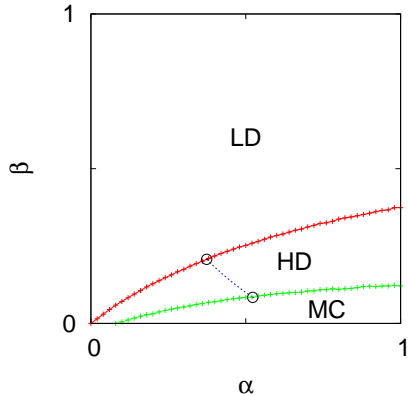}
\caption{\label{fig6} Phase diagram of the open system with 
weak (reverse) bias. The
  bulk rates are $p=2/3$, $q=1$, $u=3/4$ and $v=1/4$. The low density,
  high density and maximum current phase are denoted by LD,HD and MC, 
respectively. 
The boundaries of the high density phase have been calculated by numerical
simulations. Along the dashed line given in Eq. (\ref{product}), the
semi-infinite system has a product measure steady state. The circles denote
the exactly known points of the phase boundaries.}
\end{figure}

Next, we turn to the case of weak bias in the bulk. According to numerical
simulations, the open system has three phases, 
see the phase diagram in Fig. \ref{fig6}.  
If the exit rate exceeds a certain $\alpha$-dependent value,
$\beta_1({\alpha})$, the effective injection rate of particles at the left
boundary is not sufficiently large to force a non-vanishing current against
the bias in the bulk. The low density domain covers the entire system ($r=0$) 
and a boundary layer forms at the left boundary. The current decreases
exponentially with $L$ as given in Eq. (\ref{JSB}) with $r=0$. 
This phase is termed as low density (LD) phase.  
If $\beta$ is smaller than $\beta_1({\alpha})$ the particle injection at 
the left boundary is able to force a finite density 
$\varrho>\varrho_0$ to the bulk. 
At both ends of the system, boundary layers form but, far from the boundaries,
the profiles tend exponentially to constant values 
$\rho$ and $\pi$ in lane $A$ and $B$, respectively, 
which fulfill relation (\ref{rhopi}). 
Thus, in the bulk of a large system, the influence of boundary layers can be
neglected. To be precise, the steady state of a finite open subsystem in the
middle of the system, $[L/2-a,L/2+a]$, is of factorized form in the limit
$L\to\infty$:  
$P(\{n_i^A,n_i^B\})=\prod_{i=L/2-a}^{L/2+a}\rho^{n_i^A}(1-\rho)^{1-n_i^A}\pi^{n_i^B}(1-\pi)^{1-n_i^B}$.
As a consequence, the current is given by Eq. (\ref{fund}) in the limit 
$L\to\infty$ and, in accordance with $\varrho>\varrho_0$, it is non-vanishing: 
$J_{\infty}\equiv\lim_{L\to\infty}J(L)>0$. 
In this phase, which is termed as high density (HD) phase,
the bulk densities $\rho$ and $\pi$, as well as the current are
non-trivial functions of $\alpha$ and $\beta$. 

At the phase boundary between the LD and the HD phase 
($\beta=\beta_1({\alpha})$), 
a similar phenomenon can be observed as at the coexistence line of the 
one-lane ASEP. 
The HD and the LD phase coexist here in the sense that a delocalized anti-shock
develops in the interior of the system, which separates a HD domain on its left
hand side from a LD domain on its right hand side. 
At the left end of the system, exponentially decaying boundary layers 
develop in the profiles. 
In the bulk of the HD domain, the densities are $\rho_0$ and $\pi_0$
given in Eq. (\ref{rhostar}), which follows from that 
$J_{\infty}=0$. In the bulk of the LD domain, the 
densities are zero: $\rho=\pi=0$. 
The anti-shock has a finite width and its center performs symmetric diffusion
in the whole system.  
A symmetric random walk on a one-dimensional lattice of size $L$ 
with reflective boundaries has a steady state with a uniform 
probability $1/L$ at each site. 
For a sharp domain wall and no boundary layer at the left end this would lead 
to profiles which interpolate linearly between $\rho_0$($\pi_0$) and zero in
lane $A$($B$). For a finite system this holds only approximatively for large
$L$ and far from the boundaries: 
$\langle n^A_i\rangle\approx(1-i/L)\rho_0$, $\langle n^B_i\rangle\approx(1-i/L)\pi_0$, 
$1\ll i\ll L$. 
Nevertheless, on the macroscopic scale, i.e. introducing the rescaled variable
$x=i/L$ and performing the limit $L\to\infty$, the profiles are given as:
\be 
\langle n^A_x\rangle=(1-x)\rho_0,  
\qquad \langle n^B_x\rangle=(1-x)\pi_0
\ee    
for $0<x<1$. 
The dependence of the current on the system size $L$ can be easily obtained. 
Measuring the current at the right end of the system, it is clear that 
particles
have a significant chance to leave the system at the right boundary and
contribute to the mean current only when the domain wall is found in the
vicinity of the right boundary. The probability of this event is proportional 
to $L^{-1}$, therefore the  
current in a finite system decreases with $L$ as: 
\be
J(L)\sim L^{-1}.
\ee

In the HD phase, the current $J_{\infty}$ is determined by the boundary rates
at the left boundary and, for a fixed $\alpha$, it increases with decreasing
$\beta$. 
For given bulk rates, however, 
the current which can flow in the bulk is
bounded and has a maximal value $J_{\rm max}$, see 
the fundamental diagram in Fig. \ref{fig4}. 
Therefore, if $\alpha$ is not too small and 
$\beta$ is decreased the current attains its maximum at some $\beta=\beta_2(\alpha)>0$. 
If $\beta_2(\alpha)>0$ and $\beta<\beta_2(\alpha)$, the current is saturated, 
i.e. $J_{\infty}=J_{\rm max}$, and 
the bulk densities $\rho_{\rm max}$ and $\pi_{\rm max}$ in lane $A$ and $B$,
respectively, are given by the equation $J_{\infty}(\rho_{\rm max})=J_{\rm max}$ 
and by Eq. (\ref{rhopi}).
These densities, as well as the current are independent of the boundary rates
$\alpha$ and $\beta$. 
The density profiles tend algebraically to the bulk value with the distance
$l$ measured from the boundaries as $|\langle n^A_l\rangle-\rho|\sim
l^{-1/2}$, $|\langle n^B_l\rangle-\pi\sim l^{-1/2}$ for  $1\ll l\ll L$. 
The value of the decay exponent $1/2$ is related to that the fundamental
diagram is quadratic in the vicinity of the maximum \cite{krug}. 
After the analogous phase of the ASEP, this phase will be called 
maximum current (MC) phase. 

In general, we could not determine the phase boundaries $\beta_1(\alpha)$ and 
$\beta_2(\alpha)$ exactly apart from
certain special cases.
For a given set of bulk rates, we have performed numerical simulations for
different boundary rates $\alpha$ and $\beta$ and
estimated the phase boundaries by measuring the current in the steady state.
The results are shown in Fig. \ref{fig6}.  
The system size and simulation time was $L=300$ and $t=2\times10^6$,
respectively, in case of the upper boundary $\beta_1(\alpha)$, whereas 
$L=1000$ and $t=5\times10^6$ in case of the lower boundary $\beta_2(\alpha)$. 

A special case, where the upper boundary can be determined exactly is $p=q$. 
If vacancies(particles) in lane $B$ are regarded as particles(vacancies), 
one can see that none of
the two lanes is singled out if $\alpha=\beta$. 
The rates at the left boundary thus force a high density domain to the system in which 
$\langle n^A_i\rangle_{HD}=1-\langle n^B_i\rangle_{HD}$ holds. Here, the
subscript HD refers to that the average is restricted to configurations where
site $i$ lies in the high density domain, i.e. $i\ll l$ where $l$ is the
location of the domain wall.  
Thus, in the bulk of the high density domain $\rho=1-\pi$ and the
corresponding current is zero. Therefore the phase boundary 
curve $\beta_1(\alpha)$ coincides in this special case with the 
diagonal $\alpha=\beta$.     

Furthermore, there is a line in the phase diagram, where the semi-infinite
system has a product measure steady state. 
It is easy to see that if the densities of particle reservoirs 
$\rho=\alpha/p$ and $\pi=1-\beta/q$ are compatible in the sense that 
they fulfill 
Eq. (\ref{rhopi}), moreover, the system is in the 
HD phase, i.e. $\rho_0<\alpha/p<\rho_{\rm max}$, 
then any finite open subsystem $[1,a]$ ($a<\infty$) 
has a factorized steady state 
in the limit $L\to\infty$ with homogeneous densities $\rho$ and $\pi$ in lane
$A$ and $B$, respectively. 
Expressing the densities of boundary reservoirs with the boundary rates 
and substituting them in Eq. (\ref{rhopi}),  we obtain 
\be
\beta_{\rm fact}(\alpha)=
q\left(1+\frac{u}{v}\frac{\alpha}{p-\alpha}\right)^{-1}.
\label{product}
\ee
This curve is plotted in the phase diagram in Fig. \ref{fig6}. 
On this line, there is no boundary layer at the left boundary. 
The upper phase boundary is reached at $\alpha=p\rho_0$, i.e. 
$\beta_1(p\rho_0)=\beta_{\rm fact}(p\rho_0)$, whereas the lower phase boundary
is reached at $\alpha=p\rho_{\rm max}$, i.e. 
$\beta_2(p\rho_{\rm max})=\beta_{\rm fact}(p\rho_{\rm max})$.
This provides one exactly known point of each phase boundary curve, which are
shown in Fig. \ref{fig6}.

\section{Inhomogeneous model}

In the rest of this work, we shall investigate the steady state
of the model in the case where the system is heterogeneous,
which means that the jump rates are site-dependent.
If the bulk rates are random variables, 
one can easily imagine that there may be both regions
where particles feel a local forward bias and regions with a local
reverse bias. 
Such a situation has been studied in the disordered one-lane partially
asymmetric simple exclusion process
\cite{stanley,tripathy,krug2,jsi,barma,j2}. Note that, in the totally
asymmetric process, such a situation cannot be realized.  
The key to the understanding of such heterogeneous models 
at a phenomenological level is the steady state of a one-barrier system, 
i.e. when a homogeneous region with a
reverse bias is embedded in a homogeneous medium with a
forward bias. 

\subsection{One-barrier system}

Let us consider a periodic system of size $\mathcal{L}$, 
which is composed of a homogeneous subsystem (called barrier)  
with a reverse bias ($\Delta U>0$), which 
contains sites $i=1,2,\dots,L$ and the rest of the system (the environment),
where there is a forward bias ($\Delta U<0$).
The particle density $\varrho=N/(2\mathcal{L})$ is fixed, 
furthermore, we assume that $1\ll
L=b\mathcal{L}$, where $b\ll 1$, thus the barrier is only a small part of the
system so that a single particle has a positive 
average velocity (in a finite system).
Similar to the reverse bias regime, one must distinguish between two types of
forward bias. If $\Delta U<0$ and $\Delta \overline{U}>0$, we speak of 
strong forward bias, while if $\Delta U<0$ but $\Delta \overline{U}<0$ we
speak of weak forward bias.  
Depending on whether the bias is strong or weak at the barrier and in the
environment we have four different cases. 

\begin{figure}[h]
\includegraphics[width=0.5\linewidth]{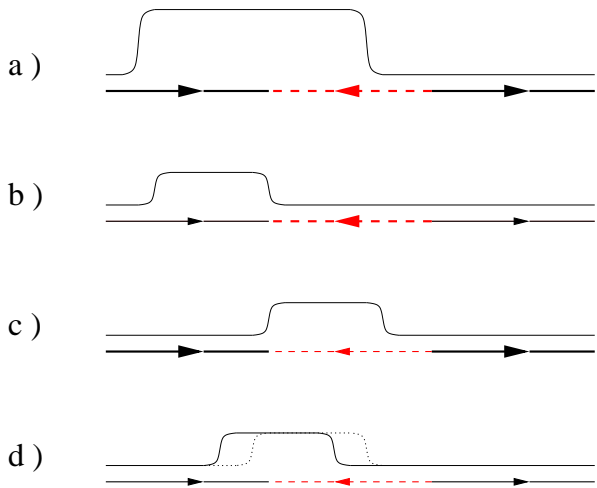}
\caption{\label{fig7} Density profiles in different one-barrier
  systems. Arrows indicate the direction of the bias and big(small)
  arrow heads correspond to strong(weak) bias.}
\end{figure}

Let us assume that the bias is strong in both parts of the system. 
In this case, almost all
particles are found in a high density cluster of length 
$l\simeq N/2=\varrho\mathcal{L}$, where $\rho=\pi=1$, see Fig. \ref{fig7}a.  
The right edge of the HD domain (the anti-shock)
lies in the barrier region and covers a fraction $r$ of this
region as given in Eq. (\ref{r}), provided the number of particles is
sufficiently large: $N/2\simeq l>2rL$. 
Focusing on the barrier region, the steady state is similar to that of
an open system with $\alpha>0$ and $\beta=0$. 
If, however, $N/2<2rL$ the HD
cluster is located symmetrically on the two sides of site $1$ 
and covers only a fraction $r=l/(2L)=N/(4L)$ of the barrier region.  
The stationary current decreases exponentially with $L$ as given
in Eq. (\ref{JSB}). 

Next, we consider the case when the bias is still strong in the
reverse bias region but it is weak in the environment.
Now, a HD cluster forms in the forward bias region, the right edge of which 
is located at site $1$, see Fig. \ref{fig7}b. 
In the HD domain, the densities $\rho_0$ and $\pi_0$ 
are given by Eq. (\ref{rhopi}) with the rates of the forward bias region. 
As particles detaching from the anti-shock region must overcome the empty
barrier region, the stationary current is exponentially vanishing, 
as given in Eq. (\ref{JSB}) with $r=0$. 

Let us consider the case when the bias is strong in the
environment but it is weak in the barrier region. 
Reversing the directions left and right, this arrangement is identical to the
previous one, except that the barrier region is limited.
Nevertheless, until  $N<2L\varrho_0$ or, equivalently, if the length $l=N/(2\varrho_0)$ of a HD cluster
with density $\varrho_0$ is less than $L$, a HD domain will form 
 in the reverse bias region as we have known in the previous case. 
This is illustrated in Fig. \ref{fig7}c.
At the left edge of the HD domain (at site $1$), a boundary layer forms
in the forward bias region, 
which effectively closes the reverse bias
region, and the situation is similar to the open system with
$\alpha=\beta=0$.
Since, the right
edge of the HD cluster does not reach the right edge of the barrier region, 
the current is exponentially vanishing as given in Eq. (\ref{JSB})
with $r=l/L$. 
The steady state is, however, different if $N>2L\varrho_0$. 
In this case, the HD domain covers the entire reverse bias region and, here, 
the density exceeds $\varrho_0$. Besides, the density of particles 
is non-vanishing also in the environment and, 
accordingly, the current is non-vanishing ($J_{\infty}>0$), as well. 
 
The most complex situation is when the bias is weak both in the barrier region
and in the environment. 
Restricting our attention to the reverse bias region, this can be
regarded as an open system of size $L$, with some effective boundary
rates $\alpha_{\rm eff}$ and $\beta_{\rm eff}$, which depend on the
bulk rates in a non-obvious way owing to the
presence of boundary layers around site $1$.  
Since both effective boundary
rates are finite, the determination of the steady state
in the general case is, in
addition to the above difficulties, as hard as the
determination of the phase diagram for an open system with weak
reverse bias.
We shall therefore restrict ourselves to the description of 
the possible steady states and omit to
give criteria in terms of the jump rates except of some special cases
when the symmetries of the model can be exploited.

For almost all points of the parameter space spanned by 
$p_f$,$q_f$,$u_f$,$v_f$,$p_r$,$q_r$,$u_r$,$v_r$, where the index
$f$($r$) refers to the forward(reverse) bias region, the steady state
is similar to those described in the previous two cases. 
That means, a high density cluster forms which is located either entirely 
in the forward bias region (Fig. \ref{fig7}b) or entirely in the barrier
region (Fig. \ref{fig7}c) and the densities are given by the
densities $\rho_0$ and $\pi_0$ of the medium which correspond to zero 
current, see Eq. (\ref{rhostar}).    
In the latter case, the current can be non-vanishing ($J_{\infty}>0$)
if the density of particles is sufficiently high.   

In addition to the above steady states, there is a zero measure set in
the parameter space for which the high density cluster becomes delocalized, 
analogous to the behavior at the phase boundary between the LD and the
HD phase in the open system (see Fig. \ref{fig7}d). 
In this case, the right edge of the high density cluster lies in the
reverse bias region and the whole cluster performs symmetric
diffusive motion so that the right edge of the cluster moves  
in the range $[1,\min\{L,l_r\}]$, where $l_r=N/(2\varrho^r_0)$ is the
length of the cluster corresponding to the zero-current density 
$\varrho^r_0$ in the reverse bias region. 
In the part of the cluster which lies in the forward (reverse) bias
region, the local densities assume the zero-current densities 
of the underlying medium, i.e. $\rho^f_0$ and $\pi^f_0$ 
($\rho^r_0$ and $\pi^r_0$). 
If $l_r/L=\varrho/(b\varrho^r_0)\ge 1$, then the current vanishes with the
size of the barrier region as
$J(L)\sim L^{-1}$, otherwise it decreases exponentially.    
The symmetric diffusive motion of the cluster results in a linearly decaying
average density profile in the barrier region.

The general condition of such a delocalized steady state is not known but we
can provide two sufficient conditions, which are related to the symmetries of
the model. 
First, if $p_f=q_r$, $q_f=p_r$, $u_f=v_r$ and $v_f=u_r$ then the two sides of
site $1$ (i.e. the forward bias region and the reverse bias region) are mapped
to each other by rotating the system by $180$ degrees about the
axis perpendicular to the plane in which it is drawn in Fig. \ref{fig1}. 
That means that the two regions are equivalent and 
the currents in the two lanes cancel, independent of the 
actual location of the HD cluster. Fluctuations of the net current lead to 
displacements of the latter and, as a consequence, it is 
delocalized in the steady state. 
Second, if $p_f=q_f=p_r=q_r$, the high density cluster is delocalized again,
independent of the lane change rates. 
This latter case is related to the symmetry already mentioned in Sec.  
\ref{oswb}. When the particles(vacancies) in lane $A$ are regarded as 
vacancies(particles) then the particles in both lanes move in the 
same direction and lane change of the original model transforms to pair annihilation
and creation. If the above condition is satisfied the lanes in the transformed
model are equivalent. Therefore, the particle currents in the two lanes of the
original model compensate each other independent of the location
of the particle cluster, which leads to delocalization.        
 Although, boundary layers develop around site $1$ in the profiles in lane $A$ and $B$, 
for the total density $\langle n^A_i\rangle_{HD}+\langle n^B_i\rangle_{HD}=1$
holds in the HD cluster due to the above symmetry.

\begin{figure}[h]
\includegraphics[width=0.5\linewidth]{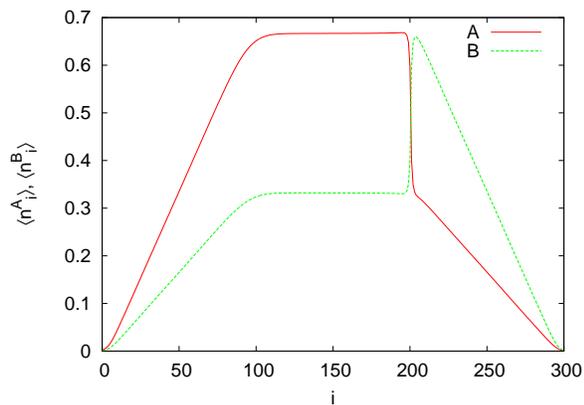}
\caption{\label{fig8} Numerically calculated density profiles in a periodic
  one-barrier system. The size of the system is $\mathcal{L}=800$, the number 
of particles is $N=100$ and the reverse bias region is located in the interval 
$[200,300]$. Transition rates are $p_f=q_f=p_r=q_r=0.5$ and
  $v_r=1-u_r=u_f=1-v_f$. The simulation time was $10^9$. Only the part of
the system is shown where the density is significantly differs from zero. In
the rest of the system, which is never reached by the high density cluster, the
density is vanishing: $\varrho\approx J/[p_f(v_f-u_f)]\sim\mathcal{L}^{-1}$.}
\end{figure}

\begin{figure}[h]
\includegraphics[width=0.5\linewidth]{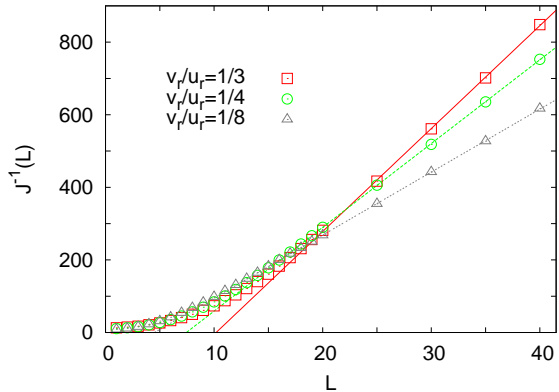}
\caption{\label{fig9} The inverse of the measured stationary current as a
  function of the size $L$ of the barrier in a periodic one-barrier system for
  different lane change rates. The
  size of the system was $\mathcal{L}=16L$ and the number of particles was
  $N=4L$. Transition rates are $p_f=q_f=p_r=q_r=0.5$ and
  $v_r=1-u_r=u_f=1-v_f$. The simulation time was $10^9$.}
\end{figure}

We have performed numerical simulations in the second case and measured the
stationary density profiles (see Fig. \ref{fig8}), as well as the 
current for different system sizes. 
The results on the current, which are
shown in Fig. \ref{fig9}, are in agreement with the law
$J(L)\sim L^{-1}$ for large system sizes.
The intercept $L^*$ of the linear asymptote 
$J^{-1}(L)\simeq s(L-L^*)$ with the $x$
axis can be interpreted as the effective width of the anti-shock region. 
As can be seen, for smaller $v_r/u_r$, the anti-shock is sharper and 
the finite size corrections to the asymptotic law are weaker. 
The slope $s$ of the asymptote depends on  $v_r/u_r$, as well, 
through the sharpness of the anti-shock.  
\subsection{Disordered system}

Next, we shall study the model in the presence of disorder, where the
transition rates are site-dependent quenched random variables. 
The general case is rather complex as, in such systems, both weak and
strong bias regions may form. 
We shall therefore restrict the quantitative analysis to two special
cases where either exclusively strong or exclusively weak bias 
regions are found in the system.

\subsubsection{Longitudinal disorder}

First, let us consider the case where the lane change rates are
homogeneous, i.e. $u_i=v_i\equiv u$ ($i=1,2\dots,L$) while the other
rates $p_i$ and $q_i$ are independent quenched random variables. 
The model with such a {\it longitudinal disorder} may describe 
traffic of molecular motors on heterogeneous tracks
containing defects where the advance of motors is hindered. 

In this case, the potential difference $\Delta U_{ij}$ 
between site $i$ and $j$($>i$) felt by a particle and the
corresponding potential difference $\Delta \overline{U}_{ij}$
of a vacancy are related as: 
\be 
\Delta U_{ij}=-\Delta \overline{U}_{ij} + 
\ln\frac{(p_j+u)(q_{j+1}+u)}{(p_i+u)(q_{i+1}+u)}.
\ee
The second term on the r.h.s. is an $O(1)$ random variable (with zero
mean), which is
negligible for large $j-i$ compared to the first term which increases 
linearly with $j-i$ in the case of a global bias and increases 
as $\sim\sqrt{j-i}$ in the absence of a global bias. 
Thus the two potentials are asymptotically perfectly
correlated: 
\be 
\Delta U_{ij}\simeq-\Delta \overline{U}_{ij}.
\label{long}
\ee
As a consequence, there are only strong (forward or reverse) bias 
regions but no weak bias regions in the case of longitudinal disorder. 
This is the case also in the disordered one-lane partially asymmetric
simple exclusion process, therefore the model behaves similar to that one 
and we can make us of the known results on that model.
  
In the following, we assume that the potential landscape is tilted, i.e. 
$[\Delta U_i]_{\rm av}<0$, where $[\cdot ]_{\rm av}$ denotes average
over the distribution of jump rates. 
A given realization of the random potential $U_i$, $i=1,2,\dots,L$ can
be regarded as a random walk of $L$ steps with step lengths $\Delta
U_i$, in the presence of a bias in the negative direction. 
The effective reverse bias regions in the potential landscape
can be identified with the ascending parts of Brownian excursions
in the positive direction \cite{jsi,j2}.  
A Brownian excursion of length $n$ is a random walk $U_i$, 
$i=0,1,\dots,n$, which satisfies that $U_n=U_0$ and 
$U_i>U_0$ for $0<i<n$.  
The amplitude of the excursion is defined as 
$U_{\max}-U_0\equiv\max_{0\le i\le n}{U_i}-U_0$.
The effective reverse bias region is identified with the part of the
excursion from $i=0$ up to the first maximum:
\be  
l\equiv\min_{0\le j\le n}\{j|U_j=U_{\max}\}.
\label{lmax}
\ee
  
\begin{figure}[h]
\includegraphics[width=0.5\linewidth]{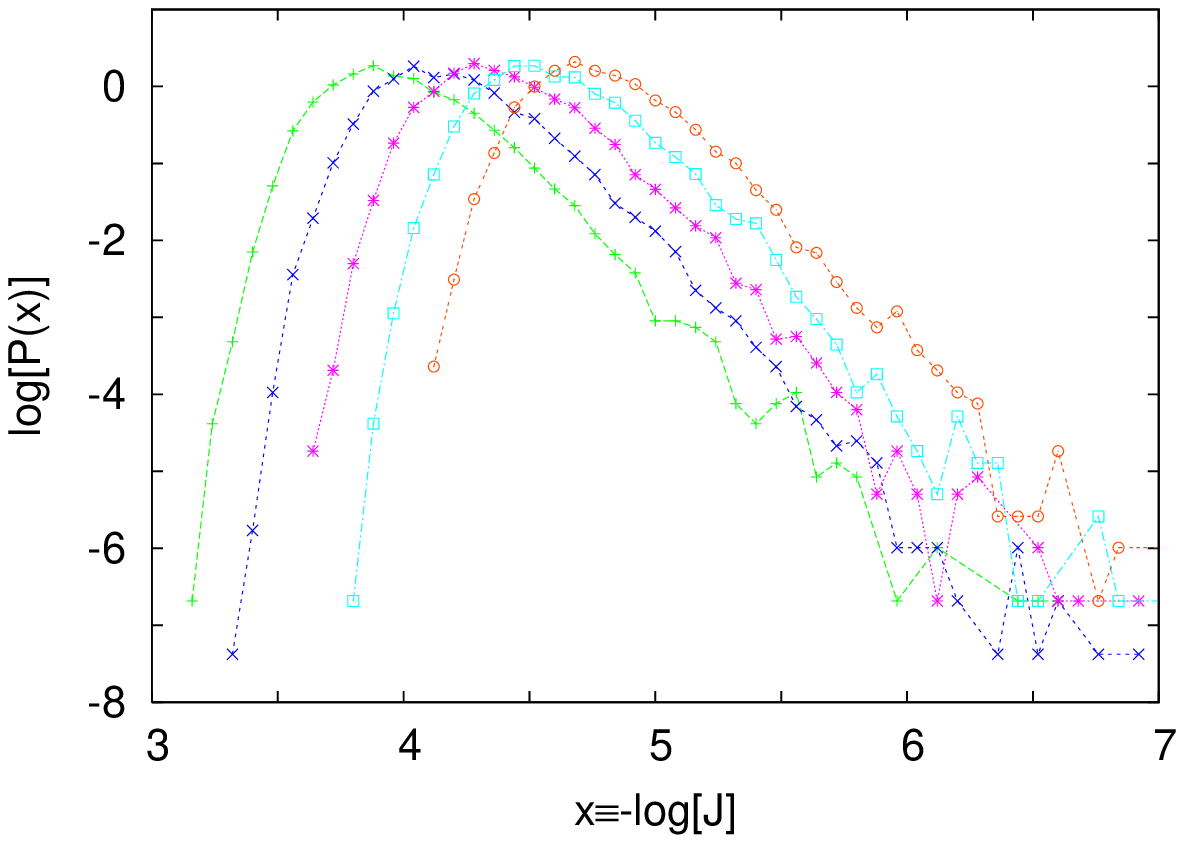}
\includegraphics[width=0.5\linewidth]{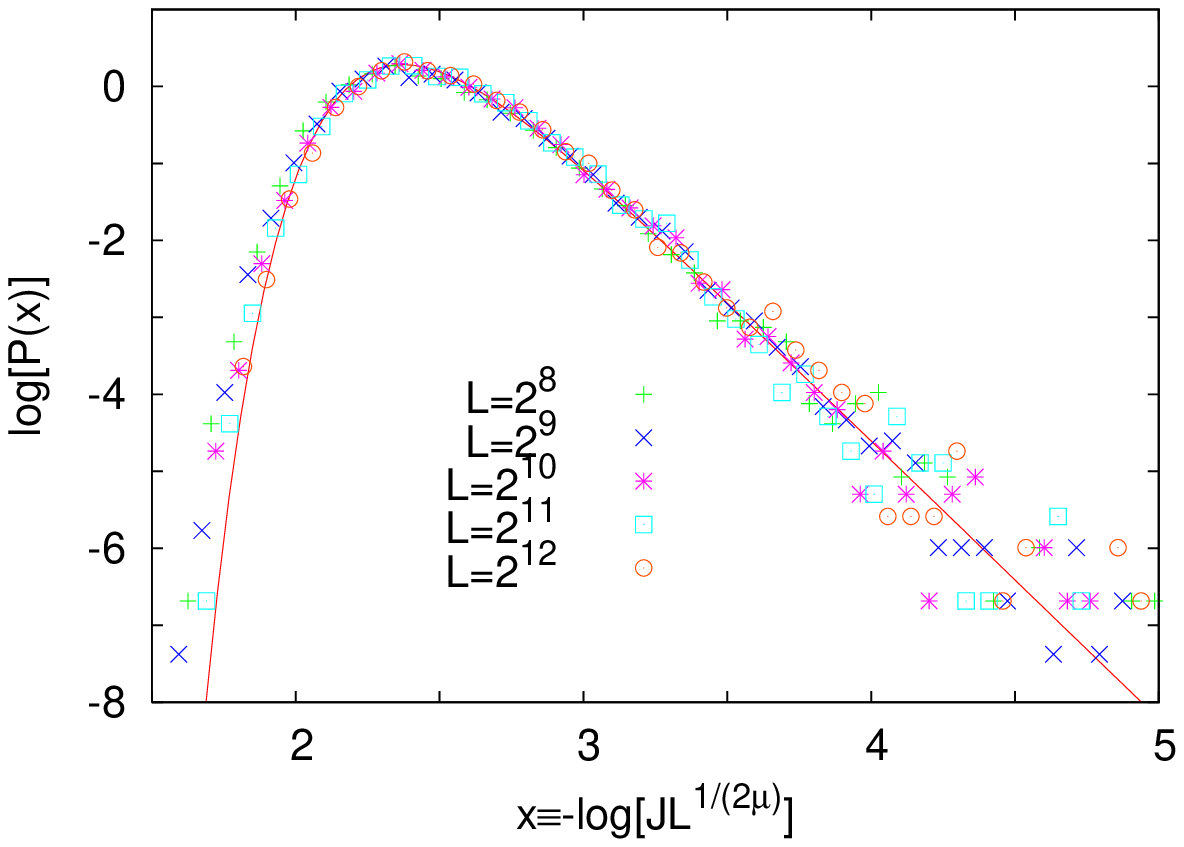}
\caption{\label{fig10} Left: Distribution of the logarithm of the current
  obtained by numerical simulations for different system sizes in case of
 the bimodal longitudinal disorder given in Eq. (\ref{bimodal}) with $\lambda=0.25$ and $c=0.2$.
Right: Scaling plot with the exponent $1/(2\mu)=0.2767288\cdots$ predicted by
the theory. The solid line is the Fr\'echet distribution given in Eq. (\ref{frechet}).}
\end{figure}

In the steady state of a finite system, the current is determined by 
the reverse bias region with the largest amplitude 
$\Delta U_{\rm MAX}$, behind which the particles accumulate and form a high
density cluster, where the density is close to $1$. 
Due to the particle-hole symmetry, which is reflected by
Eq. (\ref{long}), half-filling is realized in this maximal reverse bias region,
i.e. the anti-shock is located where the potential (measured from the
bottom of the barrier) is $\Delta U_{\rm MAX}/2$. 
The current is therefore given by $J=1/\sqrt{\tau_{\rm MAX}}$, where
 $\tau_{\rm MAX}=e^{\Delta U_{\rm MAX}}$ is the waiting time 
of a single particle in the maximal reverse bias region.
The distribution of the amplitudes of Brownian excursions decays asymptotically
exponentially (see the Appendix) or, equivalently,       
the probability density of one-particle waiting times at reverse bias
regions has an algebraic tail $p(\tau)\simeq A\tau^{-1-\mu}$. Here, the 
diffusion exponent $\mu$ is the positive root of the equation 
$[\left(\tilde q/\tilde p\right)^{\mu}]_{\rm av}=1$ in case 
of a one-lane system with independent, random 
rates $\tilde p_i$ and $\tilde q_i$ (see e.g. Ref. \cite{bouchaud}).
In terms of the jump rates of the original two-lane model, this can be written as:
\be 
\left[\left(\frac{q(p+u)}{p(q+v)}\right)^{\mu}\right]_{\rm av}=1.
\label{mugen}
\ee
According to extreme value statistical considerations \cite{galambos}, the
current, which is related to the largest one-particle waiting time,     
follows the Fr\'echet distribution  
\be      
p(\tilde J)=2\mu \tilde J^{2\mu-1}e^{-\tilde J^{2\mu}},
\label{frechet}
\ee
in terms of the scaling variable $\tilde J=cJ(L)L^{\frac{1}{2\mu}}$, 
where the constant $c$ is
related to the pre-factor $A$ of the parent distribution $p(\tau)$. 
The typical current vanishes with the system size $L$ as:
\be 
J(L)\sim L^{-1/(2\mu)}.
\ee   
We have performed numerical
simulations for the model with rates $u=0.5$, $p_i=1-q_i$ for all $i$
and rate $p_i$ was drawn from a bimodal distribution 
\be 
\rho(p)=c\delta(p-\lambda(1+\lambda)^{-1})+
(1-c)\delta(p-(1+\lambda)^{-1}),
\label{bimodal}
\ee
with $\lambda=0.25$ and $c=0.2$.  
For this bimodal randomness, the exponent $\mu$ is given by
\be 
\mu = \left[\ln (c^{-1}-1)\right]\left[
\ln\frac{\lambda+(1+\lambda)u}{\lambda+\lambda(1+\lambda)u}\right]^{-1},
\label{mu}
\ee 
which yields $1/(2\mu)=0.2767288\dots$ for $\lambda=0.25$ and $c=0.2$.
We have measured the stationary current in $10^{6}$ 
Monte Carlo steps in $10^4$ independent samples for system sizes 
$L=2^i$, $i=8,9,10,11,12$. 
The results on the distribution of the current shown in Fig. \ref{fig10} are in agreement with the
above predictions. 

\subsubsection{Transversal disorder}

Let us now consider another type of randomness, where the longitudinal
rates are homogeneous, i.e. $p_i=q_i\equiv p$ for all $i$, whereas the
transversal rates $u_i$ and $v_i$ are random variables. 
This model will be termed {\it transversal disorder} and may describe
situations where the lane change of molecular motors is biased randomly in
space due to the presence of forces transversal to the filament.
   
\begin{figure}[h]
\includegraphics[width=0.6\linewidth]{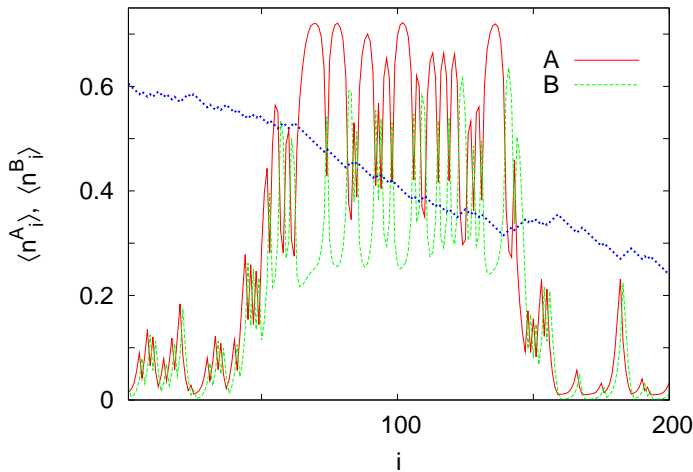}
\caption{\label{fig11} Numerically calculated density profiles in the 
steady state in the presence of transversal disorder. The system size is 
$L=200$ and the number of particles is  $N=100$. Bimodal randomness has been
used given in Eq. (\ref{bimodal2}) with $c=1/3$, $\lambda=1/8$ and  
$p=0.5$. The dotted line indicates the potential
landscape $U_i$. The longest reverse bias region is in the domain
$[139,170]$. Behind this region, a high density cluster of approximate length
$100$ can be observed, where $\langle n^A_i\rangle+\langle n^B_i\rangle\approx 1$.}
\end{figure}

In this model, the one-particle potential and one-vacancy potential are
again perfectly correlated: 
\be 
\Delta U_{i}=\Delta \overline{U}_{i}. 
\ee
Consequently, there are exclusively weak (forward or reverse) bias regions
in the system, while strong bias regions are lacking. 
Let us assume in the following that the potential landscape is tilted, 
i.e. $[\Delta U_i]_{\rm av}<0$. 
Similar to the longitudinal disorder, the environment in which the
particles move can be regarded as a set of isolated reverse bias
regions which are embedded in a medium with forward bias. 
We shall apply a phenomenological independent-barrier model 
to this system, some elements of which have already been mentioned in the 
previous section. This theory which gives exact results for a single
particle in a random environment and results which are in agreement 
with numerical results for the 
disordered partially asymmetric simple exclusion process \cite{jsi}, as well
as for the present model with longitudinal disorder. 
The system will be represented by a set of $L'=O(L)$ consecutive 
barriers, which exchange particles with each others. 
There are $N_i$ particles at the $i$th barrier and the barriers are
characterized by own current functions $I_i(N_i)$ which 
give the current in the one-barrier system with $N_i$ particles, where the
other barriers are deleted.
The current $I_i(N_i)$ is monotonically increasing with $N_i$ up to a
saturation value $I^{\rm max}_i$, i.e. 
$I_i(N_i)=I^{\rm max}_i$ if $N_i$ is greater than some $N^{\rm max}_i$ which
is characteristic of the barrier. 
In what follows, we assume that the total number of particles $N$ is
sufficiently large meaning that $\sum_iN_i^{\rm max}<N$. 
Then, in the steady state, the current is determined the barrier with the minimal
carrying capacity in the system (indexed by $m$): 
$J(L')=\min_{i}\{ I^{\rm max}_i\}\equiv I^{\rm max}_m$. 
The mean number of particles $N_i$ at other barriers is determined by 
$I_i(N_i)=J(L')$ and $N_i\le N_i^{\rm max}$ for $i\neq m$, whereas   
the number of particles at the saturated barrier is $N_m=N-\sum_{i\neq
  m}N_i>N_m^{\rm max}$. These particles form a high density cluster
behind the saturated barrier, see the density profiles 
in Fig. \ref{fig11}.
As we have seen in the previous section, the
transversal disorder corresponds to the special case, 
where the high density cluster is delocalized at a barrier.
For homogeneous barrier regions, which is a good approximation for bimodal 
randomness of the type $u_i=1-v_i$, 
\be 
\rho(u)=c\delta(u-\lambda(1+\lambda)^{-1})+
(1-c)\delta(u-(1+\lambda)^{-1})
\label{bimodal2}
\ee
in the limit $c\ll 1$, 
the saturation value of the current (when $N_i>N^{\rm max}_i$) depends
asymptotically on the extension of the barrier region $l$ as 
$[I^{\rm max}(l)]^{-1}\simeq s(l-l^*)$ where $s$ and $l^*$ are common
constants for the barriers.  
Apart from this limit, the barriers are inhomogeneous, i.e. 
they may contain smaller domains with forward bias, and 
as a first guess, they are identified with the ascending
parts of Brownian excursions of the potential $U_i$, just as for
longitudinal disorder. 
Nevertheless, for inhomogeneous barriers, the HD cluster is still delocalized,
therefore the saturation value of the current 
scales with the extension $l_i$ of the barrier 
defined in Eq. (\ref{lmax}) as $I^{\rm max}(l_i)\simeq l_i^{-1}$.
Thus, the current in a finite system is expected to scale as 
\be
J\sim l_{\rm max}^{-1}, 
\label{jlmax}
\ee
where $l_{\rm max}\equiv\max_{1\le i\le L'}\{l_i\}$ is the length of the longest reverse bias region present in the system. 

\begin{figure}[h]
\includegraphics[width=0.5\linewidth]{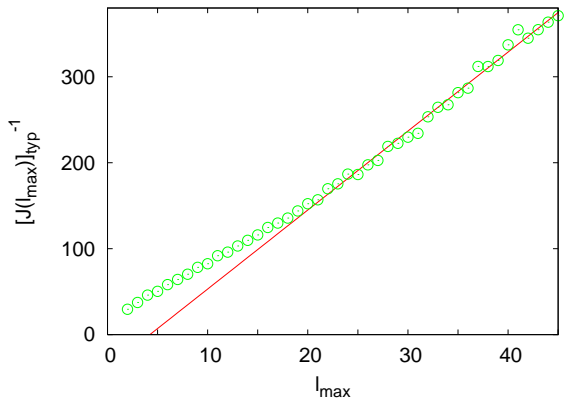}
\caption{\label{fig12} Dependence of the typical current
  on the length of the longest reverse bias region in the model with
  transversal disorder. The size of the system was $L=256$ 
and bimodal randomness in Eq. (\ref{bimodal2}) has been used 
with $c=1/3$ and $\lambda=1/8$. The number of particles was $N=128$.}
\end{figure}

We have performed numerical simulations for the model with transversal disorder
and measured the current in the steady state. We have determined in 
each sample the size $l_{\rm max}$ of the longest effective reverse bias region. 
Then the measurement was carried out for $10^5$ independent samples and 
the average of the logarithm of the current has been calculated over samples
where $l_{\rm max}$ was the same. 
The dependence of the calculated typical current 
$[J(l_{\max})]_{\rm typ}=\exp([\ln J(l_{\max})]_{\rm av})$
on $l_{\rm max}$ is shown in Fig. \ref{fig12}.  
As can be seen, the numerical results are in agreement with the law given in
Eq. (\ref{jlmax}). 
Independent of the precise definition of effective 
reverse bias regions, one may
generally conclude that their probability of occurrence must be
 exponentially small in their size $l$ 
(apart from possible corrections like in the case of continuous-time
Brownian excursions \cite{majumdar}) since the
majority of the lane change rates in such a region is of that type which
results in reverse bias (in the particular case in Eq. (\ref{bimodal2})
that with probability $c$).   
Therefore, the length of the longest effective
reverse bias region scales with the size of the system $L$ as 
$l_{\rm max}\sim \ln L$. 
Finally, we obtain that the current scales with $L$ in the case of transversal
disorder as 
\be 
J(L)\sim (\ln L)^{-1}.
\label{jln}
\ee 
We have computed the probability of amplitude $\Delta U$ of Brownian
excursions and the probability of the extension $l$ of the effective reverse
bias region which is given by the time at which the maximum is first
assumed in Brownian excursions; for the details of the calculation, see the Appendix. 
The results are plotted in
Fig. \ref{fig13}.
\begin{figure}[h]
\includegraphics[width=0.5\linewidth]{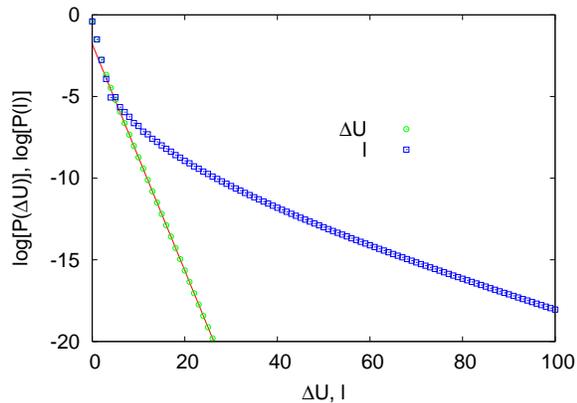}
\caption{\label{fig13} Probability of the amplitude $\Delta U$ 
(in units $\ln\frac{p+\lambda/(1+\lambda)}{p+1/(1+\lambda)}$) and the
probability of the time at which the maximum is first assumed 
in Brownian excursions 
corresponding to bimodal randomness with $c=1/3$. Latter quantity is
identified with the length $l$ of effective
reverse bias regions. The straight line is the asymptotic form given in Eq. \ref{asymp}.}
\end{figure}
The probability of amplitudes, which is the relevant variable in case of
longitudinal disorder, tends rapidly to the
asymptotical exponential from.
Contrary to this, the probability of the extension $l$ of effective reverse
bias regions, which is the important variable in case of transversal disorder, 
has strong corrections to the exponential for moderate $l$. 
Therefore the asymptotic decay in Eq. (\ref{jln}) is expected to be observed 
for rather large system sizes which are beyond the realm of 
numerical simulations.
   
\section{Summary}

We have studied in this work a two-lane model that describes bidirectional 
transport of particles interacting by hard core exclusion.
Although, the one-particle dynamics in such a two-channel environment are
qualitatively similar to those in a one-lane partially asymmetric system which 
 allows backward steps of the particle, the many particle steady-state behavior
 can be much different in the two models. 
We have investigated situations when the track contains regions where the
preferred direction of motion of particles is opposite to the global bias. 
In this case, we have found two types of steady states and have given 
the conditions of them in terms of the transition rates. 
In the case of strong reverse bias, which can be induced by making the rates
of transitions parallel to the track asymmetric, 
the steady state is similar to that of the
one-lane partially asymmetric exclusion process: a compact, localized cluster
of particles forms and the current vanishes exponentially with
the extension of the reverse bias region. 
In case of weak bias, realized by rendering the lane change rates asymmetric, 
which has no counterpart in the one-lane asymmetric
model, a qualitatively different steady state is observed. 
The cluster of particles becomes now delocalized and, as a consequence, 
the current vanishes inversely proportionally with the 
size of the barrier region. 
We have thus found that, compared to analogous one-lane systems, 
there may be physically different steady states, in which the flow
against the local external drive is facilitated by the
cooperative behavior of particles. 
It is an interesting question whether this type of delocalization can 
be observed in experiments or in living cells.  

\appendix
\section{}   

Let us consider the bimodal randomness given in
Eq. (\ref{bimodal}) and introduce 
$x_i\equiv U_i/\ln\frac{p+\lambda/(1+\lambda)}{p+1/(1+\lambda)}$.
This rescaled potential landscape $\{x_i\}$ is a random walk with steps of unit length
with probabilities $c$ and $1-c$ in the positive and negative directions, respectively.
Let us assume that the walk starts at $x_0=1$ and absorbing walls are
located at $x=0$ and at $x=X+1>1$.   
A central quantity in the computation of the statistics of the maximum is the
probability that the walker is found after $t$ steps at site $x$: 
$P_X(x,t;c)$. The probability that the walk ends at the absorbing site $X+1$
without ever having crossed the starting point is called persistence
probability and it is given as $p_{\rm pers}(X;c)=\lim_{t\to\infty}P_X(X+1,t;c)$.
This quantity can be exactly calculated \cite{ir} and reads in the present
case: 
\be
p_{\rm pers}(X;c)=(1/c-2)[(1/c-1)^{X+1}-1]^{-1}.
\ee 
The probability that the amplitude $x_{\max}$ of the walk before it crosses
the wall at $x=0$ can be written in terms of the persistence probability as
\be
{\rm Prob}(x_{\max}=x)=c[p_{\rm pers}(x-1;c)-p_{\rm pers}(x;c)]
\ee 
for $x>1$. 
For $x=0$ and $x=1$, we have ${\rm Prob}(x_{\max}=0)=1-c$ and ${\rm Prob}(x_{\max}=1)=c(1-c)$.
For large $x$, this probability decays exponentially:
\be 
{\rm Prob}(x_{\max}=x)\simeq(1-2c)^2/(1-c)(1/c-1)^{-x}.
\label{asymp}
\ee    

Next, we turn to the computation of the distribution of the time $t_{\max}$
at which the walk first reaches its maximum before it crosses the
wall at $x=0$. It is easy to see that ${\rm Prob}(t_{\max}=0)=1-c$ and 
${\rm Prob}(t_{\max}=1)=c(1-c)$.
For $t_{\max}>1$, the joint probability that the amplitude is $x_{\max}$ and it is
assumed for the first time at time $t_{max}$ is given as 
$c^2P_{x_{\max}-1}(x_{\max}-1,t_{max}-2;c)p_{\rm pers}(x_{\max};1-c)$.
From this, the marginal distribution of $t_{\max}$ is obtained as 
\be 
{\rm Prob}(t_{\max}=t)=c^2\sum_{x_{\max}=2}^{t}P_{x_{\max}-1}(x_{\max}-1,t-2;c)p_{\rm pers}(x_{\max};1-c).
\ee
The quantities $P_X(x,t;c)$ which appear in the above expression 
can be computed iteratively from those at time $t-1$ using the
initial condition $P_X(1,0;c)=1$ and $P_X(x,0;c)=0$ for $x\neq 1$.

\ack
This work has been supported by the Hungarian National Research Fund
under grant no. OTKA K75324.

\end{document}